# Dynamics of Cu$_2$O Rydberg excitons - real density matrix approach

**Karol Karpiñski and Gerard Czajkowski**

Technical University of Bydgoszcz Al. Prof. S. Kaliskiego 7 85-789 Bydgoszcz, Poland

E-mail: `czajk@pbs.edu.pl`

**Abstract.** Basing on recent reliable measurements of the life- and coherence time of Cu$_2$O Rydberg excitons, we present an analytical method which enables to describe the experimental results. The real-density matrix approach (RDMA) is applied to describe two-photon absorption in Cu$_2$O through the excitonic response to the exciting fields. This approach produces an analytical expression for the emission intensity time evolution, and can be used to explain trends in reported experimental data. Our approach takes into account the effect of the coherence between the electron-hole pair and the electromagnetic fields, the quantum beats and the life- and coherence time dependence on the applied laser power. Adding a separation of slow- and rapid dynamics components we find excellent agreement with the experimental data.

# 1. Introduction

Since the first observation of Rydberg excitons (RE) in Cu$_2$O[1] they become a subject of intensive studies (for a recent review see, for example, [3]). The research started from optical properties of REs in natural crystals, then it evaluated towards REs in low-dimensional nanostructures (see, for example, [4, 5, 6] and references therein). Both linear and nonlinear optical properties were examined. The theory, so far, is well-developed and the theoretical results agree well with experimental findings. The majority of experiments and theoretical considerations on the optical properties of Rydberg excitons was devoted to excitons created by a harmonic-time dependent wave of the type $\mathbf{E}(\mathbf{r}, t) = \mathbf{E}_0 \exp(i\mathbf{kr} - i\omega t)$ where $\mathbf{E}$ is the electric field vector, $\omega$ the frequency (circular frequency) and $\mathbf{k}$ the wave vector. Since the optical functions are represented by time averages, they remain time-independent, and the exciton (exciton population) time dependence was not discussed. Recently, the trend changed to experiments and theory on the dynamic processes [7, 8], where excitons are created by pulsed excitation. The exciton population, created within a finite time, decays then with a time constant, which is related to the so-called exciton life time.

The relevant equations on excitons created in stationary way contain an number of parameters, and most of them are known (for example electron- and hole effective masses, quantum states oscillator strengths, energy gap, dielectric constant, exciton Rydberg energy etc.) However, there is a one important parameter, which is not exactly known. It is the above mentioned exciton lifetime of a quantum state $n$ (notation $T_{2n}, \tau_n$), which corresponds to an energy $\Gamma_n = \hbar/T_{2n}$, also called the damping constant (homogeneous broadening). The coefficients $\Gamma_n$ represent dissipative processes that, in general (and we will see it below) are energy and temperature dependent [9, 10, 11]. The topic of calculation of exciton lifetime has not been covered yet. There are some estimation, as, for example $\Gamma_n \propto (n^2 - 1)/n^5$ [12] which gives the tendency $\tau_n \propto n^3$. Very often the damping $\Gamma$ was assumed as a free parameter, and was determined by fitting experimental results, thus in an indirect way.

In a recent paper [7] the authors reported on a direct measurement of the lifetime and coherence time of Cu$_2$O Rydberg excitons up to principal quantum number $n = 9$. They used a method, where a degenerate two-photon excitation (2PA) generates a one-photon emission at twice the pump energy. They used short (a few ps) laser impulses and the exciton lifetime exceeds the impulse duration. When the impulse vanishes, the excitons decay for each Rydberg state. The slope of the decay rate gives directly the lifetime $\tau$, and coherence time $\tau_{\mathrm{coh}}$, by the relation $\tau_{\mathrm{coh}} = 2\tau$. During the decay process one observes oscillations (the so-called quantum beats) due to the interference of the neighboring exciton states. In addition, in [7] the dependence of the lifetime on the laser pump power and the probe temperature, is measured. The general tendency is the decrease of the lifetime with the increase of the pump power (temperature).

Since a two photon excitation is a nonlinear process (a $\chi^{(3)}$ process), in the theoretical description of the above reported process we advance the application of the real density matrix approach (RDMA, also called coherent wave theory). This approach has been successful in describing linear and nonlinear optical properties of semiconductors in terms of Rydberg excitons. Recently it was used in the theoretical description of two photon non-degenerate absorption in silicon, in the case of stationary excitation [14]. Here we extend this approach to the case of degenerate 2PA and finite-time excitation impulses. As compared with earlier works on the application of RDMA to Rydberg physics (see, for example, [16]), the RDMA scheme is supplemented by quantum beatings equations, time-scale separation, and the dependence of exciton lifetime on applied laser power.

The paper is organized as follows. In Section 2 we recall the basic equations of the RDMA, adapted to the case of degenerate 2-photon absorption. In Section 3 we explicitly derive the formulas for linear and nonlinear susceptibility for a Cu$_2$O crystal, irradiated by a finite-time pulse. Next, in Section 4 we calculate the exciton life- and coherence times and other parameters needed for intensity calculations. In Section 5 we derive the final expression for the nonlinear time-dependent 2P emission intensity. Finally, in Section 6, we present the Fourier transforms of the theoretical emission intensity. The Appendix contains details of the presented calculation.

# 2. Two-photon absorption in RDMA formalism

## 2.1. The constitutive equations

In the RDMA approach the bulk nonlinear response will be described by a closed set of differential equations ("constitutive equations"): one for the coherent amplitude $Y(\mathbf{r}_1, \mathbf{r}_2)$ representing the exciton density related to the inter-band transition, one for the density matrix for electrons $C(\mathbf{r}_1, \mathbf{r}_2)$ (assuming a non-degenerate conduction band), and one for the density matrix for the holes in the valence band, $D(\mathbf{r}_1, \mathbf{r}_2)$. Below we will use the notation

$$Y(\mathbf{r}_1, \mathbf{r}_2) = Y_{12}, \quad \text{etc,} \qquad (1)$$

The constitutive equations have the form: the inter-band equation

$$i\hbar \frac{\partial Y_{12}}{\partial t} - H_{eh} Y_{12} = -\mathbf{M}\mathbf{E}(\mathbf{R}_{12})$$
$$+ \mathbf{E}_1 \mathbf{M}_0 C_{12} + \mathbf{E}_2 \mathbf{M}_0 D_{12} + i\hbar \left(\frac{\partial Y_{12}}{\partial t}\right)_{\text{irrev}}, \quad (2)$$

conduction band equation

$$i\hbar \frac{\partial C_{12}}{\partial t} + H_{ee} C_{12} = \mathbf{M}_0 (\mathbf{E}_1 Y_{12} - \mathbf{E}_2 Y_{21}^*)$$
$$+ i\hbar \left(\frac{\partial C_{12}}{\partial t}\right)_{\text{irrev}}, \quad (3)$$

valence band equation

$$i\hbar \frac{\partial D_{21}}{\partial t} - H_{hh} D_{21} = \mathbf{M}_0 (\mathbf{E}_2 Y_{12} - \mathbf{E}_1 Y_{21}^*)$$
$$+ i\hbar \left(\frac{\partial D_{21}}{\partial t}\right)_{\text{irrev}}, \quad (4)$$

where the operator $H_{eh}$ is the two-particles (electron and hole) effective mass Hamiltonian

$$H_{eh} = E_g - \frac{\hbar^2}{2M_{tot}} \partial_Z^2 - \frac{\hbar^2}{2M_{tot}} \boldsymbol{\nabla}_{R_\parallel}^2$$
$$- \frac{\hbar^2}{2\mu} \partial_z^2 - \frac{\hbar^2}{2\mu} \boldsymbol{\nabla}_\rho^2 + V_{eh}, \quad (5)$$

with the separation of the center-of-mass coordinate $\mathbf{R}_\parallel$ from the relative coordinate $\boldsymbol{\rho}$ on the plane $x-y$,

$$H_{ee} = -\frac{\hbar^2}{2m_e}(\boldsymbol{\nabla}_1^2 - \boldsymbol{\nabla}_2^2),$$
$$H_{hh} = -\frac{\hbar^2}{2m_h}(\boldsymbol{\nabla}_1^2 - \boldsymbol{\nabla}_2^2), \quad (6)$$

and $\mathbf{E}_{12}$ means that the wave electric field in the medium is taken in a middle point between $\mathbf{r}_1$ and $\mathbf{r}_2$: we take them at the center-of-mass

$$\mathbf{R} = \mathbf{R}_{12} = \frac{m_h \mathbf{r}_1 + m_e \mathbf{r}_2}{m_h + m_e}. \quad (7)$$

In the above formulas $m_e, m_h$ are the electron and the hole effective masses (more generally, the effective mass tensors), $M_{tot}$ is the total exciton mass, and $\mu$ the reduced mass of electron-hole pair. The smeared-out transition dipole density $\mathbf{M}(\mathbf{r})$ is related to the bi-locality of the amplitude $Y$ and describes the quantum coherence between the macroscopic electromagnetic field and the inter-band transitions (see, for example, [5]). The terms $(\partial Y/\partial t)_{\text{irrev}}$ etc. describe the irreversible dissipation and radiation decay processes due to all dephasing processes. The resulting coherent amplitude $Y_{12}$ determines the excitonic part of the polarization of the medium

$$\mathbf{P}(\mathbf{R}, t) = 2 \int d^3 r \, \mathbf{M}^*(\mathbf{r}) \text{Re} \, Y(\mathbf{R}, \mathbf{r}, t)$$
$$= \int d^3 r \mathbf{M}^*(\mathbf{r})[Y(\mathbf{R}, \mathbf{r}, t) + \text{c.c}], \quad (8)$$

where $\mathbf{r} = \mathbf{r}_1 - \mathbf{r}_2$ is the electron-hole relative coordinate. Having in mind experiments described in [7], we consider a situation where the irradiating electromagnetic field $\mathbf{E}$ includes two frequencies $\omega_1$ and $\omega_2$, and is written as

$$\mathbf{E} = \mathbf{e}_\alpha F(t) \Big[ \mathcal{E}_{01} \exp(i\mathbf{k}_1 \mathbf{R} - i\omega_1 t)$$
$$+ \mathcal{E}_{02} \exp(i\mathbf{k}_2 \mathbf{R} - i\omega_2 t) + \text{c.c.} \Big], \quad (9)$$

with the envelope function $F(t)$ (the impulse shape, the definition will be given below), and

$$|\mathbf{k}_j| = \frac{\omega_j}{c} \sqrt{\epsilon(\omega_j)} = n_j \frac{\omega_j}{c}, \quad j = 1, 2, \quad (10)$$

and $n_j$ are refractive indices at the frequencies $\omega_j$. In the following, we consider only one component $(\mathbf{A})_\alpha$ of the vectors $\mathbf{E}, \mathbf{P}$ and $\mathbf{M}$. The linear optical properties are calculated by solving the inter-band equation (2), supplemented by the corresponding Maxwell equation, where the polarization (8) acts as a source. For computing the nonlinear optical properties we use the entire set of constitutive equations (2-4). Although finding a general solution of the equations is challenging, in special situations a solution can be found. For example, if one assumes that the matrices $Y, C$ and $D$ can be expanded in powers of the electric field $\mathbf{E}$, an iterative procedure can be used. In general, solving for $Y$, $C$ and $D$ in the context of two-photon absorption depends on the relation between the incoming frequencies $\omega_1, \omega_2$ (and thus energies $\hbar\omega_1$, $\hbar\omega_2$) and the fundamental gap energy $E_g$, which enters as a parameter in the electron-hole Hamiltonian. We consider the case, when $\hbar\omega_1 + \hbar\omega_2 \leq E_g$, the excitation of discrete excitonic states is possible.

*2.2. The iterative solution of constitutive equations*

Our goal is to derive expressions for the two-photon absorption coefficients. These can be obtained from the third-order nonlinear susceptibility, which, in turn, can be determined via an iterative procedure within the context of the RDMA. To apply the iteration procedure we assume, that the relevant quantities $P, Y, C$ and $D$ can be expanded in series of the input electric field amplitudes. The considered inter-band transition is composed of two steps. The first step starts from the initial state $E_i$ (here the maximum $E_v$ of the valence band) to an intermediate (virtual) state $\beta$, the second is from the state $\beta$ to the final state $E_f$ (the minimum of the conduction band $E_c$). The first step in the iteration consists of solving the equation (2), which at this stage takes on the form

$$i\frac{\partial Y^{(1)}}{\partial t} - H_{eh}^{(j)} Y^{(1)} = -\mathbf{M}\mathbf{E} + i\hbar \left(\frac{\partial Y^{(1)}}{\partial t}\right)_{\text{irrev}}, \quad (11)$$

where **E** has the form (9). The Hamiltonians $H_{j,eh}$ are defined with regard to the two steps of transition described above:

$$H_{1,eh} = E_\beta - E_v + H_r, \quad (12)$$

$$H_{2,eh} = E_c - E_\beta + H_r, \quad (13)$$

$$H_r = \left(-\frac{\hbar^2}{2}\right) \nabla_r \left(\underline{\underline{\mu}}\right)^{-1} \nabla_r + V(\mathbf{r}), \quad (14)$$

where $\underline{\underline{\mu}}$ is the exciton reduced mass tensor for given pair of bands $C$ and $V$, and $V(\mathbf{r})$ the dielectrically screened electron-hole Coulomb potential, $E_\beta$ is the energy of the virtual band, and $E_c, E_v$ correspond to extreme od valence and conduction bands.

When considering the time evolution of the exciton population, generate by the impulse (9), two time scales should be accounted for. The processes of exciton creation are rapid processes, on the femtosecond scale. The time characterizing the decline process of the population is proportional to the lifetime $\tau$, which is of the order $3\tau < 20$ ps, i.e. is a slow process. Therefore the amplitudes $Y_j^{(1)}$ will consist of two parts

$$Y_j^{(1)}(\mathbf{r},t) = Y_{\text{slow}}^{(1)}(t) Y_{0j}(\mathbf{r}) e^{-i\omega_j t}. \quad (15)$$

For the irreversible part we assume the simple form

$$\left(\frac{\partial Y^{(1)}}{\partial t}\right)_{\text{irrev}} = -\frac{1}{\tau} Y^{(1)}. \quad (16)$$

As follows from experiments, the decline of excitonic absorption is governed by the rule $\propto \exp(-t/\tau_{n\ell m})$. To obtain this behavior, we distinguish slow and fast changes (in other words, two time scales) in the excitonic amplitudes $Y$ and matrices $C, D$. With regard to (9), we put

$$Y^{(1)} = \tilde{Y} e^{-i\omega t} \quad (17)$$

into (11), obtaining

$$\frac{d\tilde{Y}}{dt} = -i\omega \tilde{Y} e^{-i\omega t} + e^{-i\omega t} \frac{d\tilde{Y}}{dt} + e^{-i\omega t} i\Omega_{eh} \tilde{Y}$$

$$- e^{-i\omega t} \left(\frac{d\tilde{Y}}{dt}\right)_{\text{irrev}} = \frac{i}{2\hbar} F(t) e^{-i\omega t} M(\mathbf{r}), \quad (18)$$

$$i(\Omega_{eh} - \omega)\tilde{Y} + \frac{d\tilde{Y}}{dt} - \left(\frac{d\tilde{Y}}{dt}\right)_{\text{irrev}} = \frac{i}{2\hbar} F(t) M(\mathbf{r}).$$

Taking the 2 steps in 2-photon process, we have two pairs of equations

$$i\hbar \left[i\omega_1 + \frac{1}{\tau}\right] Y_{1+}^{(1)} - H_{eh}^{(1)} Y_{1+}^{(1)} = -M\mathcal{E}_{01}^*, \quad (19)$$

$$i\hbar \left[-i\omega_1 + \frac{1}{\tau}\right] Y_{1-}^{(1)} - H_{eh}^{(1)} Y_{1-}^{(1)} = -M\mathcal{E}_{01},$$

with similar equations for $Y_{2\pm}^{(1)}$, $M$ is the relevant component of the transition dipole density. Now we separate the slow and rapid processes, by putting

$$\tilde{Y}^{(1)} = Y_0^{(1)}(\mathbf{r}) Y_{\text{slow}}^{(1)}. \quad (20)$$

In what follows we consider only the resonant part. The parts $Y_0^{(j)}$ correspond to rapid processes, and have the form

$$Y_{01-}^{(1)}(\mathbf{r}) = \mathcal{E}_{01} \sum_{n\ell m} \frac{c_{n\ell m} \varphi_{n\ell m}(\mathbf{r})}{E_\beta + E_{n\ell m} - \hbar\omega_1 - i\hbar\tau_{n\ell m}^{-1}},$$

$$Y_{02-}^{(1)}(\mathbf{r}) = \mathcal{E}_{02} \sum_{n\ell m} \frac{c_{n\ell m} \varphi_{n\ell m}(\mathbf{r})}{E_c - E_\beta + E_{n\ell m} - \hbar\omega_2 - i\hbar\tau_{n\ell m}^{-1}},$$

$$E_\beta = E_v + \hbar\omega_1, \quad (21)$$

where $E_v, E_c$ are the extreme of the valence and conduction bands, respectively. The coefficients $c_{n\ell m}$ and eigenfunctions are defined as follows

$$c_{n\ell m} = \int d^3 r M(\mathbf{r}) \varphi_{n\ell m}(\mathbf{r}),$$

$$\varphi_{n\ell m} = R_{n\ell}(r) Y_{\ell m}(\theta, \phi), \quad (22)$$

$Y_{\ell m}$ are the spherical harmonics, $R_{n\ell}$ are the radial functions of a corresponding hydrogen atom-like Schrödinger equation,

$$\varphi_{n\ell m}(\mathbf{r}) = R_{n\ell m}(r) Y_{\ell m}(\theta, \phi), \quad (23)$$

where

$$R_{n\ell m}(r) = N_{n\ell} \left(\frac{2\eta_{\ell m} r}{na^*}\right)^\ell$$

$$\times M\left(-n + \ell + 1; 2\ell + 2; \frac{2\eta_{\ell m} r}{na^*}\right) e^{-\eta_{\ell m} r/na^*}, \quad (24)$$

$$N_{n\ell} = \frac{1}{(2\ell + 1)!} \left[\frac{(n+\ell)!}{2n(n-\ell-1)!}\right]^{1/2} \left(\frac{2}{n}\right)^{3/2}.$$

In the above expression $M(a, b; z)$ is the confluent hypergeometric function. The energy eigenvalues related to the eigenfunctions (23) have the form

$$E_{n\ell m} = -\frac{\eta_{\ell m}^2}{n^2} R^*, \quad (25)$$

where

$$n = 1, 2, ..., \quad \ell = 0, 1, 2, ...n - 1,$$
$$m = -\ell, -\ell + 1, ..., +\ell, \quad (26)$$

$R^*$ is the effective exciton Rydberg energy for the exciton

$$R^* = \frac{\mu_\| e^4}{2(4\pi\epsilon_0 \sqrt{\epsilon_\| \epsilon_z})^2 \hbar^2}, \quad (27)$$

the anisotropy parameter $\gamma$ is defined as

$$\gamma = \frac{\mu_\| H_\|}{\mu_z \epsilon_z}, \quad (28)$$

$\mu_\parallel$ and $\mu_z$ are the heavy hole exciton reduced masses in the $x-y$ plane and in the $z$-direction, respectively,

$$\frac{1}{\mu_\parallel} = \frac{1}{m_{e\parallel}} + \frac{1}{m_{h\parallel}},$$
$$\frac{1}{\mu_z} = \frac{1}{m_{ez}} + \frac{1}{m_{hz}}. \quad (29)$$

$\epsilon_0$ is the vacuum dielectric constant, and $\epsilon_\parallel, \epsilon_z$ are relative dielectric tensor elements. The quantity $\eta_{\ell m}(\gamma)$ is given by the following expression [15]

$$\eta_{\ell m}(\gamma) = \int_0^{2\pi} d\phi \int_0^\pi \frac{|Y_{\ell m}|^2 \sin\theta\, d\theta}{\sqrt{\sin^2\theta + \gamma\cos^2\theta}}. \quad (30)$$

With regard to (25), the excitonic resonance energies result from equation

$$E_{Tn\ell m} = E_g + E_{n\ell m}. \quad (31)$$

The slow part $Y^{(1)}_{slow}$ satisfies the equation

$$\frac{dY^{(1)}_{slow}}{dt} - \left(\frac{dY^{(1)}_{slow}}{dt}\right)_{irrev} = F(t), \quad (32)$$

as previously, $F(t)$ describes the exciting impulse profile, and

$$\left(\frac{dY^{(1)}_{slow}}{dt}\right)_{irrev} = -\sigma Y^{(1)}_{slow}, \quad (33)$$

where, to simplify the notation, we put $\sigma = 1/\tau_{n\ell m}$. Equation (32) can be solved by means of Green's function, satisfying the equation

$$\frac{dG(t,t')}{dt} + \sigma G(t,t') = -\delta(t-t'), \quad (34)$$

with the solution

$$G(t,t') = \frac{1}{2\sigma} e^{-\sigma|t-t'|}. \quad (35)$$

Thus, the solution of (32) has the form

$$Y^{(1)}_{slow}(t) = \int_{-\infty}^\infty F(t')G(t,t')\, dt'. \quad (36)$$

As follows from the definition of $\sigma$, $Y^{(1)}_{slow}$ will be labelled by quantum numbers $n, \ell, m$.

### 2.3. Coherence time and exciton life time

The values of the above mentioned exciton life time can be taken from experiments (see [7] Table I.) We can also estimate them, using the relation between the coherence time and the exciton life time. Coherence time is the time duration over which the medium

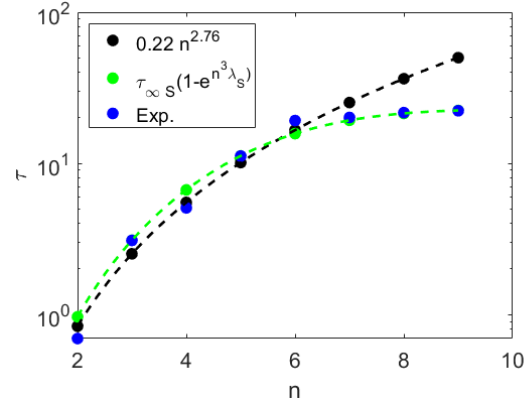

**Figure 1.** Exciton lifetimes, directly measured S states lifetimes [7] (blue dots), in good agreement with the $n^{2.7}$ scaling (black dashes), up to $n = 6,(42)$, for $n > 6$ good agreement with scaling (43) for S excitons

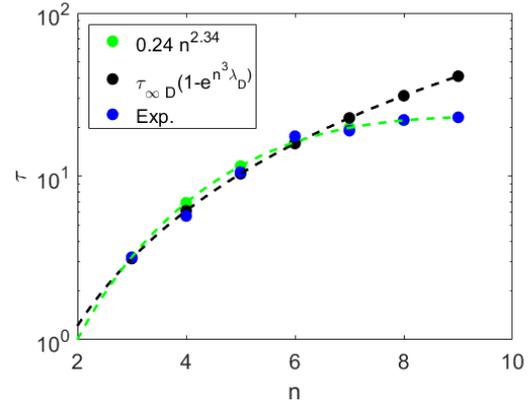

**Figure 2.** The same as Figure 1, for $D_2$ states, for $n \leq 6$ in good agreement with $n^3$ scaling, for $n > 6$ good agreement with (43) for D excitons

impulse response is considered to be not varying. The coherence time, which we denote as $\tau_{coh}$, is calculated by dividing the coherence length by the phase velocity of light in a medium; approximately given by

$$\tau_{coh} = \frac{1}{\Delta\nu} \approx \frac{\lambda^2}{c\,\Delta\lambda}, \quad (37)$$

where $\lambda$ is the central wavelength of the source, $\Delta\nu$ and $\Delta\lambda$ is the spectral width of the source in units of frequency and wavelength respectively, and $c$ is the speed of light in vacuum. Identifying the variation of impulse response with energy difference between neighboring exciton states we obtain (without specification to S or D states)

$$\tau_{coh,n} = \frac{\delta\hbar}{E_{n+1} - E_n}, \quad (38)$$

where $\delta$ is a certain parameter. Below we take the values $\delta_S = 19.7$ and $\delta_{D1} = 19.73, \delta_{D2} = 18.72$, for S

and D1,D2 excitons, respectively, which give a fairly good agreement of theoretical and experimental values of exciton life times. If the energies are given in meV, we obtain the result in picoseconds

$$\tau_{\text{coh},n} = \frac{2\delta\hbar}{E_{n+1} - E_n}. \tag{39}$$

On the other hand, we have the relation

$$\tau_{coh,n} = 2\tau_n, \tag{40}$$

$\tau_n$ being exciton lifetimes. By equations (39, 40) we obtain the lifetimes

$$\tau_{n00} = \frac{\delta_S \hbar}{E_{n+1,00} - E_{n00}} = \frac{\delta_S \hbar n^2(n+1)^2}{R^* \eta_{00}^2 (2n+1)},$$

$$\tau_{n22} = \frac{\delta_D \hbar}{E_{n+1,22} - E_{n22}} = \frac{\delta_D \hbar n^2(n+1)^2}{R^* \eta_{22}^2 (2n+1)}. \tag{41}$$

The results for the S states $2 \leq n \leq 9$, and D states $3 \leq n \leq 9$ are given in Tables A1 and A2. We observe a fairly good agreement with the experimental data for the states $n \leq 7$. In this interval the lifetimes scale approximately as

$$\tau_{n00} = \tau_{0S} \times n^{2.7}, \quad \text{S excitons}$$
$$\tau_{n22,1} = \tau_{0D1} \times n^{2.7}, \quad \text{D1 excitons}, \tag{42}$$
$$\tau_{n22,2} = \tau_{0D2} \times n^3, \quad \text{D2 excitons},$$

where $\tau_{0S} = 0.13$, $\tau_{0D1} = 0.15$, $\tau_{0D2} = 0.11$. The formula for D2 exciton is identical as the formula for the case of P excitons in Cu$_2$O used in [12]. The expressions (42) correspond to the scaling with quantum defects (see [7]). As observed in experiments, the values of lifetimes for $n > 7$ arrive at a plateau. For higher states states we observe a slow increase, tending to a value $\tau_\infty$. The scaling in this interval can be described by the formula

$$\tau_{n00} = \tau_{nS} = \tau_{\infty S}\left(1 - e^{-n^3 \lambda_S}\right), \quad \text{S excitons},$$
$$\tau_{nD1,2} = \tau_{\infty D1,2}\left((1 - e^{-n^3 \lambda_{D1,2}}\right), \quad \text{D1,D2 excitons}. \tag{43}$$

The results, for

$$\tau_{\infty S} = 22.89 \text{ ps}, \quad \lambda_S = 5.36 \times 10^{-3}, \tag{44}$$
$$\tau_{\infty D1} = 23.65 \text{ ps}, \quad \lambda_{D1} = 5.386 \times 10^{-3},$$
$$\tau_{\infty D2} = 23.14 \text{ ps}, \quad \lambda_{D2} = 6.6 \times 10^{-3}, \tag{45}$$

are given in Tables A1 and A1. The above data include the splitting into D1 and D2 excitons. The switching between the increasing of $\tau$ (those decreasing of intensity) can be explained as follows (see, for example, [17]). As we have shown above, the absorption spectrum for energies below the energy gap consists of a series of lines at energy values (31) with intensities decreasing, roughly speaking, as $1/n^3$ (those exciton lifetimes increasing, see (42)). For high values of the quantum number $n$ (i.e. for energies just below the gap), the density of exciton states per unit energy is

$$D(E) = 2\frac{\partial n}{\partial E} = \frac{2n^3}{R^*} \tag{46}$$

and, since the density decreases as $1/n^3$, we obtain a finite limit in the absorption coefficient, represented above by the quantity $\tau_\infty$, for energies near $E_g$.

### 2.4. Quantum beats

In experimental intensity curves shown in Ref. [7] some oscillations, also called quantum beats, around the smooth shape determined by a function $F(t)$ are observed. There are several types of oscillations, with different frequencies with different periodicity and amplitudes. As we show below, the oscillations are due to interference between quantum mechanical waves, generated by neighboring states, in the region of an exciton state $n$ (S or D). Consider, for example, states $|300\rangle, |322\rangle$, and $|400\rangle$, denoted further by 1,2, and 3. The state of the system is described by three waves

$$\Psi(\mathbf{r},t) = a_1\psi_1(\mathbf{r})\exp(-i\Omega_1 t) + a_2\psi_2(r)\exp(i\Omega_2 t)$$
$$+ a_3\psi_1(\mathbf{r})\exp(i\Omega_3 t), \tag{47}$$

where $a_1, a_2, a_3$ are real numbers such that

$$a_1^2 + a_2^2 + a_3^2 = 1, \tag{48}$$

$\psi_i(r)$, $i = 1,2,3$ are real valued functions, and

$$\Omega_1 = \frac{E_1}{\hbar}, \quad \Omega_2 = \frac{E_2}{\hbar}, \quad \Omega_3 = \frac{E_3}{\hbar}. \tag{49}$$

The density of probability, that the system is in a non-stationary state described by (47), is given by

$$\rho(\mathbf{r},t) = a_1^2\psi_1^2(\mathbf{r}) + a_2^2\psi_2^2(\mathbf{r}) + a_3^2\psi_3^2(\mathbf{r})$$
$$+ 2a_1a_2\psi_1(\mathbf{r})\psi_2(\mathbf{r})\cos(\Omega_{12}t) \tag{50}$$
$$+ 2a_1a_3\psi_1(\mathbf{r})\psi_3(\mathbf{r})\cos(\Omega_{13}t)$$
$$+ 2a_2a_3\psi_2(\mathbf{r})\psi_3(\mathbf{r})\cos(\Omega_{23}t),$$

where

$$\Omega_{ij} = \frac{E_i - E_j}{\hbar}. \tag{51}$$

The quantum beats are attributed to transitions between neighboring S and D states $|n00\rangle, |n22\rangle$, $|(n+1,00\rangle$. Their periodicity is given by the equations

$$\tau_{qbS,i} = \left|\frac{2\pi\hbar}{E_{n00} - E_{n+1,00}}\right|,$$
$$\tau_{qbD,i} = \left|\frac{2\pi\hbar}{E_{n22} - E_{n+1,22}}\right|, \tag{52}$$
$$\tau_{qbSD,i} = \left|\frac{2\pi\hbar}{E_{i00} - E_{n22}}\right|, \quad i = n, \ n+1.$$

The impact of quantum beats on the exciton spectra

## 3. Linear and nonlinear susceptibility

### 3.1. Linear susceptibility

The above consideration allow to calculate the total excitonic optical response in the 2 photon absorption process. This will be obtained by means of linear and nonlinear excitonic amplitudes $Y^{(1)}_{j-}(t)$ and $Y^{(3)}_{j-}(t)$. The linear amplitudes have the form

$$Y^{(1)}_{1-}(\mathbf{r},t) = \mathcal{E}_{01} \sum_{n\ell m} \frac{c_{n\ell m} \varphi_{n\ell m}(\mathbf{r})}{E_\beta - \hbar\omega_1 + E_{n\ell m} + \mathrm{i}\hbar/\tau_{n\ell m}}$$
$$\times \psi_{osc,n\ell m}(t), \qquad (55)$$

with analogous formula for $Y^{(1)}_{2-}(t)$. We use the notation

$$\bigl(G(t,t')F(t')\bigr) = \int_{-\infty}^{\infty} G(t,t')F(t')\mathrm{d}t'. \qquad (56)$$

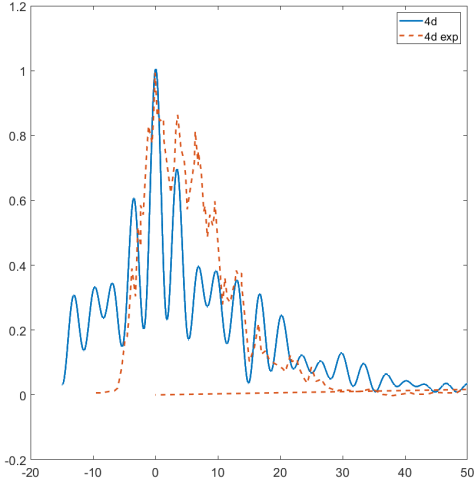

**Figure 3.** Linear time dependent emission, showing quantum beats for 4D2 excitons, the experimental curve from [7]

The solutions for $Y^{(1)}_{j\pm}(t)$ determined above allow the calculation of the linear polarization being a function of time $t$

$$P^{(1)}(\omega_1,\omega_2;t) = \int \mathrm{d}^3 r \left[Y^{(1)}_{1-} + Y^{(1)*}_{1+}\right] M^*(\mathbf{r})$$
$$+ \int \mathrm{d}^3 r \left[Y^{(1)}_{2-} + Y^{(1)*}_{2+}\right] M^*(\mathbf{r}) \qquad (57)$$
$$= \epsilon_0 \chi^{(1)}(\omega_1;t)\mathcal{E}_{01} + \epsilon_0 \chi^{(1)}(\omega_2;t)\mathcal{E}_{02},$$

where

$$\chi^{(1)}(\omega_1;t) = \frac{2}{\epsilon_0} \sum_{n\ell m} \Biggl\{ \psi_{osc,n}(t)\bigl(G_{n\ell m}(t,t')F(t')\bigr)$$
$$\times \frac{|c_{n\ell m}|^2 (E_{n\ell m} + E_\beta)}{(E_\beta + E_{n\ell m})^2 - \hbar^2(\omega_1 + \mathrm{i}/\tau_{n\ell m})^2} \Biggr\},$$
$$\qquad (58)$$
$$\chi^{(1)}(\omega_2;t) = \frac{2}{\epsilon_0} \sum_{n\ell m} \Biggl\{ \psi_{osc,n}(t)\bigl(G_{n\ell m}(t,t')F(t')\bigr)$$
$$\times \frac{|c_{n\ell m}|^2 (E_{n\ell m} + E_g)}{(E_g + E_{n\ell m})^2 - \hbar^2[(\omega_1+\omega_2) + \mathrm{i}/\tau_{n\ell m}]^2} \Biggr\}.$$

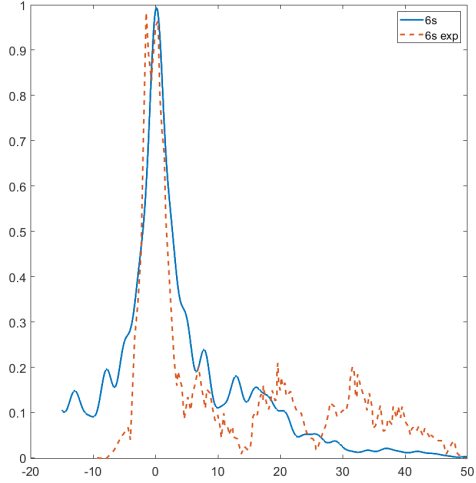

**Figure 4.** Linear time dependent emission, showing quantum beats) for 6S excitons

will be described by expressions of the type (50). The over-all expression for the oscillating terms will be taken in the form [7]

$$\psi_{osc,n}(t) = \frac{Q_n}{2}\left[1 + \sum_j q_{jn} \cos(\Omega_{jn} t)\right], \qquad (53)$$

where $j$ runs over the inter-exciton states transitions starting with $n\mathrm{S}(n\mathrm{D})$ accounted for. The quantum beats amplitudes $q_{jn}$ are proportional to the quadrupole transitions moments $Q_{jn}$, given in [8],

$$q_{jn} = \frac{Q_{jn}}{Q_n}, \quad Q_n = \sum_j Q_{jn}. \qquad (54)$$

### 3.2. Nonlinear susceptibility

In order to obtain the nonlinear response, the solutions for $Y^{(1)}_{j,pm}(t)$ are inserted as a source term in the conduction band equation (3) and the valence band equation (4). Note that each of these equations depend on the electromagnetic field. If the irreversible terms are well defined, the equations (3,4) can be solved, and this second step of the iteration yields expressions for the density matrices $C$ and $D$. We use for the irreversible terms a linear relaxation time approximation

$$\left(\frac{\partial C}{\partial t}\right)_{\mathrm{irrev}} \qquad (59)$$
$$= -\frac{1}{T_1}\left[C(\mathbf{X},\mathbf{r},t) - f_{0e}(\mathbf{r})C(\mathbf{X},\mathbf{r}=0,t)\right] - \frac{C(0)}{T_1},$$

where $\mathbf{X} = (\mathbf{r}_1+\mathbf{r}_2)/2$, $T_1$ is the carrier relaxation time, and $f_{0e}, f_{0h}$ are normalized Boltzmann distributions for electrons and holes, respectively,

$$f_{0e}(\mathbf{r}) = \int d^3q f_{0e}(\mathbf{q}) e^{-i\mathbf{qr}} \qquad (60)$$
$$= \exp\left(-\frac{m_{e\parallel}k_B\mathcal{T}}{2\hbar^2}\rho^2 - \frac{m_{ez}k_B\mathcal{T}}{2\hbar^2}z^2\right),$$

where $k_B$ is the Boltzmann constant, and $\mathcal{T}$ the temperature. The same type of expression holds for the holes. From (59) one obtains the solutions in the form

$$C(\mathbf{r},t) = -\frac{i}{\hbar}T_1 f_{0e}(\mathbf{r}) J_C(0),$$
$$D(\mathbf{r},t) = -\frac{i}{\hbar}T_1 f_{0h}(\mathbf{r}) J_D(0), \qquad (61)$$

where we present an approximation following from the assumption $T_1 \gg T_2$. The sources terms $J_C, J_D$ are defined as

$$J_C = M_0\mathcal{E}(Y_{12} - Y_{12}^*)$$
$$= 2iM_0 F(t)|\mathcal{E}_{01}|^2[\text{Im } g(-\omega_1,\mathbf{r}) + \text{Im } g(\omega_1,\mathbf{r})]$$
$$+ 2iM_0 F(t)|\mathcal{E}_{02}|^2[\text{Im } g(-\omega_2,\mathbf{r}) + \text{Im} g(\omega_2,\mathbf{r})], \qquad (62)$$

where

$$g(-\omega_1, \mathbf{r}) = \sum_{n\ell m} \frac{c_{n\ell m}\varphi_{n\ell m}(\mathbf{r})}{E_\beta + E_{n\ell m} - \hbar\omega_1 - i\hbar/\tau_{n\ell m}}$$
$$\times \bigl(G(t,t')F(t')\bigr)\psi_{osc,n\ell m} \qquad (63)$$

$$g(-\omega_2, \mathbf{r}) = \sum_{n\ell m} \frac{c_{n\ell m}\varphi_{n\ell m}(\mathbf{r})}{E_g + E_{n\ell m} - \hbar(\omega_1+\omega_2) - i\hbar/\tau_{n\ell m}}$$
$$\times \frac{\tau_{n\ell m}}{2}\bigl(G(t,t')F(t')\bigr)\psi_{osc,n\ell m},$$

and $J_c = J_D$. The above expressions can then, in turn, be used as source term for equation (2), which can which can be solved to obtain expressions for $Y_{1,2\pm}^{(3)}$ in the final step of the iteration. Now the equations for the third order coherent amplitudes $Y_{j\pm}^{(3)}$ have the form

$$i\hbar\left(i\omega_1 + \frac{1}{\tau}\right) Y_{1+}^{(3)} - H_{1,eh} Y_{1+}^{(3)}$$
$$= M_0(E^*C + E^*D) = E^*(\mathbf{R},t)\tilde{J}, \qquad (64)$$

$$i\hbar\left(-i\omega_1 + \frac{1}{\tau}\right) Y_{1-}^{(3)} - H_{eh} Y_{1-}^{(3)}$$
$$= M_0(EC + ED) = E\mathbf{R},t)\tilde{J},$$

with similar equations for $Y_{2\pm}^{(3)}$, where

$$\tilde{J} = -\frac{i}{\hbar} M_0 \left[T_1 J_C(0) f_{0e}(\mathbf{r}) + T_1 J_V(0) f_{0h}(\mathbf{r})\right]. \qquad (65)$$

Equations (64) will be solved by using the same method as in the case of linear amplitudes $Y^{(1)}$. We put $Y^{(3)}$ into the form

$$Y^{(3)} = Y_{slow}^{(3)} Y_0^{(3)}(\mathbf{r}). \qquad (66)$$

In analogy to definition (21), the stationary part of $Y^{(3)}$ will be assumed in the form

$$Y_{01-}^{(3)}(\mathbf{r}) = \mathcal{E}_{01} \sum_{\nu\lambda\mu} \frac{d_{\nu\lambda\mu,1}\varphi_{\nu\lambda\mu}(\mathbf{r})}{E_\beta + E_{\nu\lambda\mu} - \hbar\omega_1 - i\hbar/\tau_{\nu\lambda\mu}},$$

$$Y_{02-}^{(3)}(\mathbf{r}) = \mathcal{E}_{02} \sum_{\nu\lambda\mu} \frac{d_{\nu\lambda\mu,2}\varphi_{\nu\lambda\mu}(\mathbf{r})}{E_c - E_\beta + E_{\nu\lambda\mu} - \hbar\omega_2 - i\hbar/\tau_{\nu\lambda\mu}},$$

$$E_\beta = E_v + \hbar\omega_1. \qquad (67)$$

If neglecting the anti-resonant terms, we obtain

$$d_{\lambda\mu\nu,1} = \frac{iM_0 T_1}{\hbar}[\langle\varphi_{\lambda\mu\nu}|f_{0e}\rangle + \langle\varphi_{\lambda\mu\nu}|f_{0h}\rangle]2(E_{\nu\lambda\mu}+E_\beta)J_c(0)$$
$$= \frac{iM_0 T_1}{\hbar}[\langle\varphi_{\lambda\mu\nu}|f_{0e}\rangle + \langle\varphi_{\lambda\mu\nu}|f_{0h}\rangle]2(E_{\nu\lambda\mu}+E_\beta)$$
$$\times 2iM_0|\mathcal{E}_{01}|^2 \sum_{n\ell m} \frac{\hbar}{\tau_{n\ell m}} \frac{c_{n\ell m}\varphi_{n\ell m}(0)}{(E_\beta + E_{n\ell m})^2 - \hbar^2(\omega_1 + i/\tau_{n\ell m})^2},$$
$$= -M_0^2(A_{\lambda\mu\nu}+B_{\lambda\mu\nu})(E_{\nu\lambda\mu}+E_\beta)$$
$$\times |\mathcal{E}_{01}|^2 \sum_{n\ell m} \frac{T_1}{\tau_{n\ell m}} \frac{c_{n\ell m}\varphi_{n\ell m}(0)}{(E_\beta + E_{n\ell m})^2 - \hbar^2(\omega_1 + i/\tau_{n\ell m})^2}, \qquad (68)$$

$$d_{\mu\nu\lambda,2} = -4M_0^2(A_{\lambda\mu\nu}+B_{\lambda\mu\nu})(E_{\nu\lambda\mu}+E_g)$$
$$\times |\mathcal{E}_{02}|^2 \sum_{n\ell m} \frac{T_1}{\tau_{n\ell m}} \frac{c_{n\ell m}\varphi_{n\ell m}(0)}{(E_g + E_{n\ell m})^2 - \hbar^2(\omega_1+\omega_2 + i/\tau_{n\ell m})^2},$$

where

$$A_{\lambda\mu\nu} = \langle\varphi_{\lambda\mu\nu}|f_{0e}\rangle,$$
$$B_{\lambda\mu\nu} = \langle\varphi_{\lambda\mu\nu}|f_{0h}\rangle. \qquad (69)$$

The slow part $Y_{slow}^{(3)}$ satisfies the equation

$$\frac{dY_{slow}^{(3)}}{dt} - \left(\frac{dY_{slow}^{(3)}}{dt}\right)_{irrev} = F(t)\bigl(G(t,t')F(t')\bigr),$$

$$Y_{slow}^{(3)} = \bigl(G(t,t'')F(t'')\bigr)\bigl(G(t'',t')F(t')\bigr).$$

From $Y_{1\pm}^{(3)}(t), Y_{2\pm}^{(3)}(t)$ one finds the third order polarization according to

$$P^{(3)}(\omega_1,\omega_2;t) = \int d^3r \left[Y_{1-}^{(3)} + Y_{1+}^{(3)*}\right] M^*(\mathbf{r})$$
$$+ \int d^3r \left[Y_{2-}^{(3)}(t) + Y_{2+}^{(3)*}(t)\right] M^*(\mathbf{r}) \qquad (70)$$
$$= \epsilon_0 \chi^{(3)}(\omega_1,t)|\mathcal{E}_{01}|^2\mathcal{E}_{01} + \epsilon_0 \chi^{(3)}(\omega_2,t)|\mathcal{E}_{02}|^2\mathcal{E}_{02}.$$

where the nonlinear susceptibilities have the form

$$\chi^{(3)}(\omega_1,t) \qquad (71)$$
$$= -\frac{2M_0^2}{\epsilon_0} \sum_{\lambda\mu\nu} \frac{c_{\lambda\mu\nu}(A_{\lambda\mu\nu}+B_{\lambda\mu\nu})(E_{\lambda\mu\nu}+E_\beta)}{(E_\beta + E_{\lambda\mu\nu} - \hbar\omega_1)^2 + (\hbar/\tau_{\lambda\mu\nu})^2}$$
$$\times \sum_{n\ell m} \frac{T_1}{\tau_{n\ell m}} \frac{c_{n\ell m}\varphi_{n\ell m}(0)}{(E_\beta + E_{n\ell m})^2 - \hbar^2(\omega_1 + i/\tau_{n\ell m})^2}$$
$$\times \bigl(G_{\lambda\mu\nu}(t,t'')F(t'')\bigr)\bigl(G_{n\ell m}(t'',t')F(t')\bigr)\psi_{osc,\lambda\mu\nu}^2(t),$$

$$\chi^{(3)}(\omega_2, t) \qquad (72)$$
$$= -\frac{2M_0^2}{\epsilon_0} \sum_{\lambda\mu\nu} \frac{c_{\lambda\mu\nu}(A_{\lambda\mu\nu} + B_{\lambda\mu\nu})(E_{\nu\lambda\mu} + E_g)}{[E_g + E_{\lambda\mu\nu} - (\hbar\omega_1 + \hbar\omega_2)]^2 + (\hbar/\tau_{\lambda\mu\nu})^2}$$
$$\times \sum_{n\ell m} \frac{T_1}{\tau_{n\ell m}} \frac{c_{n\ell m}\varphi_{n\ell m}(0)}{(E_g + E_{n\ell m})^2 - \hbar^2(\omega_1 + \omega_2 + \mathrm{i}/\tau_{n\ell m})^2}$$
$$\times \big(G_{\lambda\mu\nu}(t,t'')F(t'')\big)\big(G_{n\ell m}(t'',t')F(t')\big)\psi^2_{osc,n\ell m}(t).$$

The formulas (71),(71) are the basic formulas, which describe the properties of 2PA excitons, excited by finite-time laser pulses.

## 4. Results on exciton lifetimes and other parameters

### 4.1. Exciton lifetimes

We have computed the emission dynamics of Rydberg excitons in Cu$_2$O created by two-photon absorption. We used the Cu$_2$O band parameters, collected in Table 1. The first step in the calculations is to establish the lifetimes of excitons for various exciton states, where we considered the even states S and D. Those lifetimes were measured in ([7]) and we formulated equations from which the lifetimes can be calculated, for various intervals of the quantum number $n$. The results are shown in Table A1, and illustrated in Figures 1 and 2. The mentioned above two features of the dependence on the state number, the increase of the lifetime for $n \leq 7$, and a transition to plateau for $n > 7$, are visible.

**Table 1.** Band parameter values for Cu$_2$O from [15], energies in meV, masses in free electron mass $m_0$, lengths in nm, $\Delta_{LT}$ from [19]

| Parameter | Value |
|---|---|
| $E_g$ | 2172.08 |
| $R^*$ | 86.96 |
| $\Delta_{LTS}$ | $5 \times 10^{-2}$ |
| $m_e$ | 0.99 |
| $m_{h\parallel}$ | 0.5587 |
| $m_{hz}$ | 1.99 |
| $\mu_\parallel$ | 0.3597 |
| $\mu_z$ | 0.672 |
| $a^*$ | 1.1 |

### 4.2. Quantum beatings periodicity

To explain the oscillations around the envelope describing the excitonic emission delay, the periodicity of these oscillations (quantum beatings) should be calculated. We use the equations (42) and (52). The experimental findings and results of the calculations are presented in Tables A3 and A4.

### 4.3. The dependence of lifetimes on laser power and temperature

The exciton life time is related to the damping constant $\Gamma$, being also the linewidth, by the relation

$$\tau = \frac{\hbar}{\Gamma}. \qquad (73)$$

The quantity $\Gamma$ depends on temperature by relation [10]

$$\Gamma(\mathcal{T}) = \Gamma_0 + \gamma_{AC}\mathcal{T} + \gamma_{LO}\left[\exp(\hbar\omega_{LO}/k_B\mathcal{T}) - 1\right]^{-1}, \qquad (74)$$

where the term $\gamma_{AC}\mathcal{T}$ is due to exciton scattering with acoustic phonons, and term nonlinear in temperature is due to interaction with LO phonons. The coefficients $\gamma_{AC}$ and $\gamma_{LO}$ represent the strength of the exciton-acoustic phonon interaction, and exciton-LO phonon interaction, respectively, while the term $\Gamma_0$ is the low-temperature limit of the line width. As can be seen from (73,74),

$$\frac{\mathrm{d}\tau}{\mathrm{d}\mathcal{T}} < 0, \qquad (75)$$

which means that the life time decreases with the increasing temperature. By given parameters $\gamma_{AC}, \gamma_{LO}$ one can derive an analytical expression for the dependence $\tau(\mathcal{T})$. The values of coherence times for increasing power can be obtained by the relation

$$\tau_{coh,n} = \tau_0 \exp\left(nb_1 - n^2 b_2 - W b_3\right). \qquad (76)$$

The coefficients $\tau_0, b_1, b_2, b_3$ ($\tau_0$ in ps, $b_3$ in 1/mW)

$$b_1 = 0.98, \qquad b_2 = 0.04,$$
$$b_3 = 6.6 \times 10^{-3}, \qquad \tau_0 = 0.394, \qquad (77)$$

were obtained from fitting experimental results from [7]. The dependence of exciton lifetimes on state number and laser power can be compared with the dependence on temperature and state number. Having in mind the experiments [7], we consider the S states $n = 5, 7$ and calculate the lifetimes as function of temperature for the given states. We use the formulas

$$\tau_n(\mathcal{T}) = \mathcal{T}_0 \exp\left(n\gamma_1 - n^2\gamma_2 - \gamma_{3n}\mathcal{T}\right),$$
$$\gamma_{35} = 0.0188, \quad \gamma_{37} = 0.0376, \quad \mathcal{T}_0 \approx 3 \times 10^{-3}, \qquad (78)$$
$$\gamma_1 = 2.592, \qquad \gamma_2 = 0.192.$$

The results are presented in Table A6.

### 4.4. Exciton oscillator strength

The oscillator strengths for the exciton state $|n\ell m\rangle$ are defined as

$$f_{n\ell m} = \left|\int d^3 r \mathbf{M}(\mathbf{r})\varphi_{n\ell m}(\mathbf{r})\right|^2. \qquad (79)$$

For further discussion we must define the shape of the transition dipole density $M$. Since we consider only one component, we take a scalar function. The shape of $M$ depends on the considered exciton symmetry. For simplicity, the S exciton function $M$ is assumed in the form

$$M(r) = \frac{M_{01}}{4\pi r_0}\delta(r-r_0), \quad (80)$$

where $r_0$ is the so-called coherence radius [15], and the coefficient $M_{01}$ is related to the longitudinal-transversal splitting energy. An example of the latter is the integrated strength of dipole moment $M_{01}$ for S excitons, obtained from the equation

$$\frac{\Delta_{LTS}}{R^*} = 2\frac{2\mu_\|}{\epsilon_0\epsilon_b\pi a^*\hbar^2}M_{01}^2. \quad (81)$$

The results for S states

$$f_{n00} = \frac{\eta_{00}^3 n(n-\eta_{00}r_0/a^*)^{2(n-1)}}{(n+\eta_{00}r_0/a^*)^{2(n+1)}}, \quad (82)$$

are displayed in Table A7, for the appropriate $\eta_{00}$ value. For comparison, we show the oscillator strengths for the isotropic case ($\eta_{00} = 1$, $r_0 \ll a^*$) ("pure"), which obey the scaling law $f_{n00} \propto 1/n^3$. The results for $\eta_{00} \neq 1$ show a deviation from this law

$$\frac{f_{n00}}{f_{100}} \approx \frac{1}{n^{3.14}}, \quad n=2,3,... \quad (83)$$

For the D excitons we use the transition dipole moment in the form[15]

$$M(r) = M_{02}\frac{Y_{22}}{r^2 r_0}e^{-r/r_0}. \quad (84)$$

For the considered $|n22\rangle$ states the eigenfunction can be put into the form

$$\varphi_{n22}(\mathbf{r}) = N_{n2}Y_{22}(\theta,\phi)\left(\frac{2\eta_{22}}{n}\frac{r}{a^*}\right)^2\exp\left(-\frac{\eta_{22}r}{na^*}\right), \quad (85)$$

where $N_{n2}$ are the normalization constants, which enter into the definition of $f_{n22}$ oscillator strengths

$$f_{n22} = M_{02}^2 N_{n2}^2 \left(\frac{2\eta_{22}\rho_0}{n}\right)^4 \exp(-2\eta_{22}\rho_0/n). \quad (86)$$

Their values are listed in Tables A7 and A8. As in the case of S excitons, we show the "pure" values of $f_{n22}$, obtained for $\eta_{22} = 1$, and given by the relation formula

$$f_{n22} = \left(\frac{2}{n}\right)^7\left(\frac{1}{5!}\right)^2\frac{(n+2)!}{2n(n-3)!} \times \text{const},$$
$$\text{const} = M_{02}^2\rho_0^4 \times 10^{-4}. \quad (87)$$

*4.5. Dependence of A and B on the laser power*

The coefficients $A_{n\ell m}, B_{n\ell m}$, defined in equations (69), depend on the temperature by definitions of $\lambda$'s

$$\lambda_{th,e} = \left(\frac{\hbar^2}{m_e k_B \mathcal{T}}\right)^{1/2}, \quad \lambda_{th,h} = \left(\frac{\hbar^2}{m_h k_B \mathcal{T}}\right)^{1/2}. \quad (88)$$

They coefficients for hole are determined for the appropriate masses ($\|$ or $z$), and have the form

$$\begin{aligned}
\bar{\lambda}_{th,e} &= \frac{\lambda_{th,e}}{a^*} = \frac{1}{a^*}\left(\frac{2\mu_\| \hbar^2}{m_e(2\mu_\|)k_B\mathcal{T}}\right)^{1/2} \\
&= \left(2\frac{\mu_\| R^*}{m_e k_B \mathcal{T}}\right)^{1/2}, \\
\bar{\lambda}_{th,hz} &= \frac{\lambda_{th,hz}}{a^*} = \frac{1}{a^*}\left(\frac{2\mu_\| \hbar^2}{m_{hz}(2\mu_\|)k_B\mathcal{T}}\right)^{1/2} \\
&= \left(2\frac{\mu_\| R^*}{m_{hz} k_B \mathcal{T}}\right)^{1/2}, \quad (89)\\
\bar{\lambda}_{th,h\|} &= \frac{\lambda_{th,h\|}}{a^*} = \frac{1}{a^*}\left(\frac{2\mu_\| \hbar^2}{m_{h\|}(2\mu_\|)k_B\mathcal{T}}\right)^{1/2} \\
&= \left(2\frac{\mu_\| R^*}{m_{h\|} k_B \mathcal{T}}\right)^{1/2}.
\end{aligned}$$

The electron effective mass is assumed to be isotropic. As stated in [7], the dependence of coherence and life times on the laser power is equivalent to a dependence on temperature. Therefore we must calculate $\lambda$'s (89) for various temperatures and, correspondingly, for various laser powers. Since we will use the quantities $A, B$ in calculations of the excitonic emission of exciton 5S state, we will focus the attention on the quantities $A_{500}, B_{500}$. Their calculated values as function of temperature/laser power are displayed in Tables A9 and A10. We observe that these quantities are decreasing with the increasing temperaturs/laser power.

## 5. Nonlinear time dependent 2P-absorption coefficient

Having linear and nonlinear susceptibilities, we obtain the total susceptibility by the relation

$$\begin{aligned}
\chi(\omega_1+\omega_2;t) &= \chi^{(1)}(\omega_1,t) + \chi^{(1)}(\omega_2,t) \\
&+ \chi^{(3)}(\omega_1,t)|\mathcal{E}_{01}|^2 + \chi^{(3)}(\omega_2,t)|\mathcal{E}_{02}|^2.
\end{aligned} \quad (90)$$

Since the contributions of terms related to $\omega_1,\omega_2$, where we take $\omega_1 = \omega_2$, are equal, we denote $\omega_1+\omega_2 = \omega$. The imaginary part of (90) defines the nonlinear absorption coefficient

$$\begin{aligned}
\alpha_{tot} &= \alpha^{(1)}(\omega) + \alpha^{(3)}(\omega)|E_{prop}|^2, \\
\alpha^{(1)} &= \frac{\hbar\omega}{\hbar c}\text{Im}\,\chi^{(1)}, \quad (91)\\
\alpha^{(3)} &= \frac{\hbar\omega}{\hbar c}\text{Im}\,\chi^{(3)}.
\end{aligned}$$

where $|E_{prop}|^2 = |\mathcal{E}_{01}|^2 = |\mathcal{E}_{02}|^2$. Having in mind the experiments[7], we will consider the absorption for a given exciton state $n\ell m$. Using (81) and (58), we obtain for S-states

$$\alpha_{n00}^{(1)}(t) = \alpha_{n00}^{(1)'}(G_n(t,t')F(t'))\psi_{osc,n00}(t), \quad (92)$$

where
$$\alpha_{n00}^{(1)'} = \frac{\hbar\omega}{\hbar c}\sqrt{\epsilon_b}\Delta_{LTS}\frac{f_{n00}\Gamma_{n00}}{2E_{n00}^2}. \qquad (93)$$

For the nonlinear contribution we obtain
$$\chi_n^{(3)}(\omega,t) = -\frac{2M_0^2}{\epsilon_0}\frac{c_n(A_n+B_n)(E_n+E_g)}{E_n^2}$$
$$\times \frac{T_1}{\tau_n}\frac{c_n\varphi_n(0)(E_n+i\Gamma_n)}{2E_g(E_n^2+\Gamma_n^2)}$$
$$\times \big(G_n(t,t'')F(t'')\big)\big(G_n(t'',t')F(t')\big)\psi_{osc,n}^2(t). \qquad (94)$$

Similar expression holds for the D-exciton contribution. Absorption coefficient $\alpha_{n00}^{(3)}$, obtained from the imaginary part of (94), has the form
$$\alpha_n^{(3)} = -\alpha_n^{(1)'}\chi_n^{(3)'}|E_{prop}|^2$$
$$\times \big(G_n(t,t'')F(t'')\big)\big(G_n(t'',t')F(t')\big)\psi_{0sc,n}^2(t), \qquad (95)$$
$$\chi_n^{(3)'} = \epsilon_0\epsilon_b\Delta_{LTS}a^{*3}\frac{1}{E_n^2+4\Gamma_n^2}\frac{T_1}{\tau_n}\varphi_n(\rho_0)(A_n+B_n).$$

We use the relation
$$|E_{prop}|^2 = 6\times 10^8 W, \quad [W]=mW, \qquad (96)$$

appropriate for $20\,\mu$m laser spot diameter, and separating the parameters appropriate for Cu$_2$O, obtaining for $n$S states
$$\alpha_n^{(3)} = -\alpha_n^{(1)'}\bigg\{\left(\frac{t_1}{\tau_n}\right)[\varphi_n(\rho_0)(A_n+B_n)]$$
$$\times 2.745\times 10^{-8}\left(\frac{n}{\eta_{00}}\right)^4\times W \qquad (97)$$
$$\times \big(G_n(t,t'')F(t'')\big)\big(G_n(t'',t')F(t')\big)\psi_{osc,n}^2(t)\bigg\},$$

where
$$t_1 = 10^{-3}T_1.$$

In calculations we use a Gaussian shaped normalized pulse
$$F(t) = F_{max}\frac{1}{\tau_p\sqrt{2\pi}}\exp\left(-\frac{t^2}{2\tau_p^2}\right), \qquad (98)$$

where $\tau_p$ is the pulse temporal duration. Using this shape we calculate the expressions $G_n(t,t')F(t')$ and, in a lowest approximation, we obtain the time dependence of the normalized emission in the form
$$I_n(t,W) = N_n\exp\left(-\frac{|t|}{\tau_n}\right)\frac{\psi_{n,osc}(t)}{Q_n} \qquad (99)$$
$$\times \big\{1-\chi_n^{(3)'}We^{b_3W}[\varphi_n(\rho_0)(A_n+B_n)]\psi_{osc,n}(t)\big\},$$

where $N_n$ are normalization constants,
$$\chi_n^{(3)'} = \left(\frac{t_1}{\tau_n'}\right)\left(\frac{n}{\eta_{00}}\right)^4 2.475\times 10^{-8}. \qquad (100)$$

The theoretical results are illustrated in Figure 4 for 5S state and three values of the applied laser power. We observe an exponential decay, governed by the exponent $-1/\tau_5$, and quantum beats. The details of calculation are given in Appendix.

## 6. Fourier transform

Having an analytical form for the oscillating functions $\psi_{n,osc}(t)$ one can easily calculate the Fourier transform of these functions. Using the standard definition
$$F(\xi) = \frac{1}{\sqrt{2\pi}}\int_{-\infty}^{\infty}f(x)e^{i\xi x}dx,$$

and substitute $t\to x,\quad \xi\to\omega$ we obtain
$$f(x) = \sum_j q_{jn}e^{-a_n|x|}\cos(\Omega_{jn}x),$$
$$a_n = \frac{1}{\tau_n},$$

where $j$ runs over the considered transitions. Then
$$F_n(\omega) \propto \sum_j \frac{q_{jn}}{a_n^2+(\Omega_{jn}+\omega)^2}. \qquad (101)$$

The above expression will be applied for the cases of 4D and 6S excitons. In the case of 4D excitons we use the frequencies corresponding to the transitions $n_1S\to n_2D$, where $n_1=3,4,5$, and $n_2=4,1;4,2;5,1;5,2$, which are displayed in Tables A3 and A4. With the above data we obtain the expression for the Fourier transform for 4D
$$F_{D4} = 28.2\bigg[1+\frac{0.0476}{1+7000(\nu-0.06)^2}$$
$$+\frac{0.32}{1+7000(\nu-0.207)^2}+\frac{0.397}{1+7000(\nu-0.28)^2}$$
$$+\frac{0.105}{1+7000(\nu-0.365)^2}+\frac{0.132}{1+7000(\nu-0.295)^2}\bigg].$$

For the 6S excitons we obtain
$$F_{S6} = 14\bigg[\frac{0.5}{1+3947(\nu-0.09)^2}+\frac{0.345}{1+3947(\nu-0.09)^2}$$
$$+\frac{0.079}{1+3947(\nu-0.134)^2}+\frac{0.06}{1+3947(\nu-0.178)^2}$$
$$+\frac{0.007}{1+3947(\nu-0.188)^2}\bigg]. \qquad (102)$$

The corresponding Fourier transforms are presented in Figures 6 and 7. In both cases the factors were chosen to be proportional to the transition quadrupole moments for $n1S\to n2D$ [8]. The numbers 28.2 in (102) and 14 in (102) give the fit to the transform obtained numerically from experimental curve. We considered the quadrupole moments for the transitions $n_1S\to n_2D_2$ states, due to the symmetry properties of $D_2$. The state $D_1$ is not a pure D state, but reveals a mixing with nearest P state (see [18]). Therefore the quadrupole moment for the transition $n_1S\to n_2D_1$ has a smaller value than the moment $n_1S\to n_2D_2$. We assumed the rate 0.8, which corresponds to the experimental observations. For the D4, 1 → D4, 2 we used the quadrupole moment for the $P4 \to S4$ transition, due to the above described mixing of states. We observe the property mentioned in [7], that the oscillations are typically dominated by a main frequency accompanied by few others of smaller amplitude. It is due to the structure of amplitudes $q_{jn}$ (54), where the major contribution comes from the state transitions $nS \to nD2$ for $n \leq 5$ and $nS \to nD1$ for $n \geq 6$, see also Tables I, II [8].

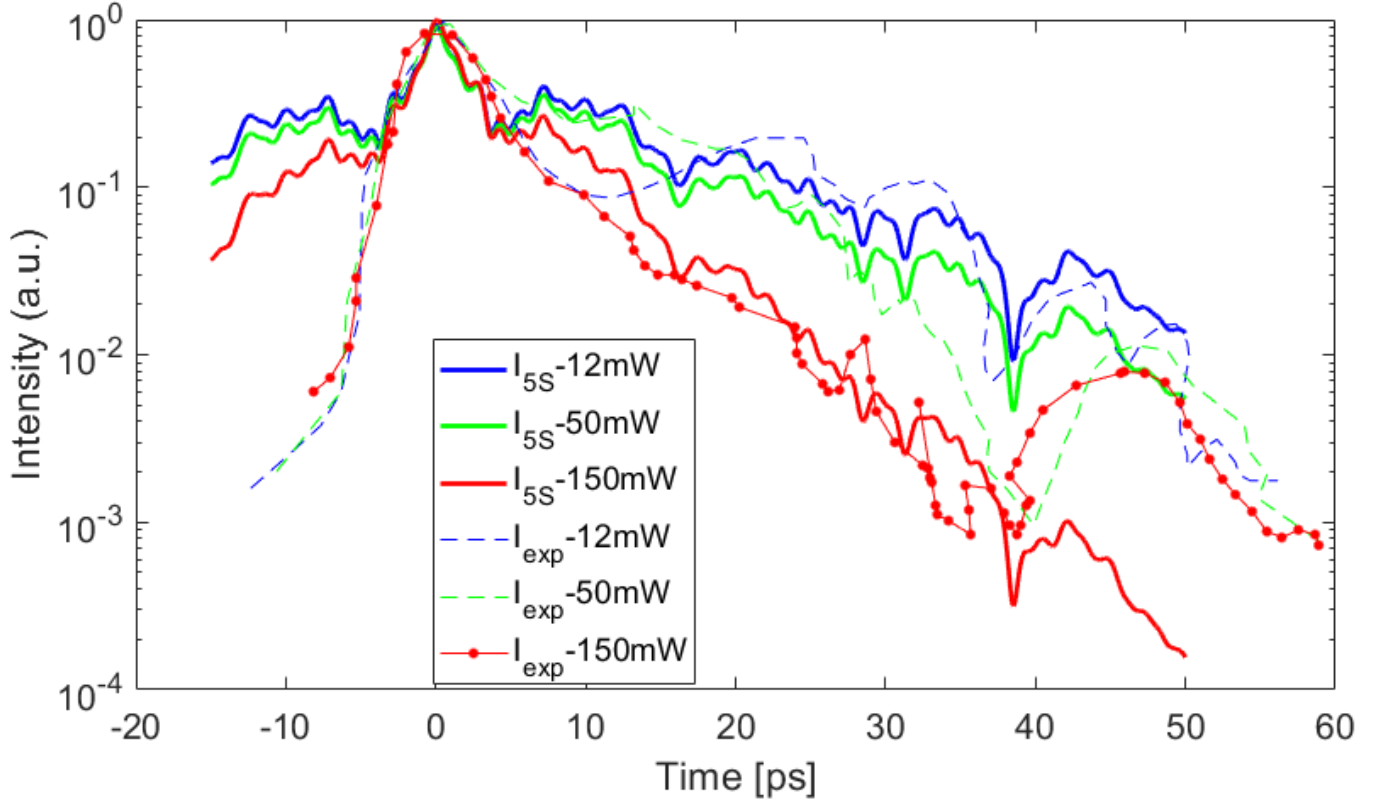

**Figure 5.** Linear time dependent emission for the state 5S and three applied laser powers

## 7. Conclusions

This study contains derivation of analytical expressions describing the direct measurements of lifetimes of the even series of Rydberg excitons in Cu$_2$O. We obtained formulas which show the power scaling laws for excitons S, D1, and D2 for the lower exciton states ($n \leq 6$) and a saturation effect for $n > 6$ states, explaining origins of this effect. Using the real density matrix approach, we derived formulas for the time dependence of linear and nonlinear susceptibility, under action of pulsed excitation. Both susceptibilities show the effect of quantum beats. The time evolution shows an exponential decay, governed by the exciton life time. In addition, we described the lifetimes dependence on the applied laser power and the crystal temperature. Having analytical expression for the nonlinear susceptibility $\chi^{(3)}$, we can calculate the nonlinear refraction index $n_2(W)$, which enables the study of Kerr-type nonlinearities [16].

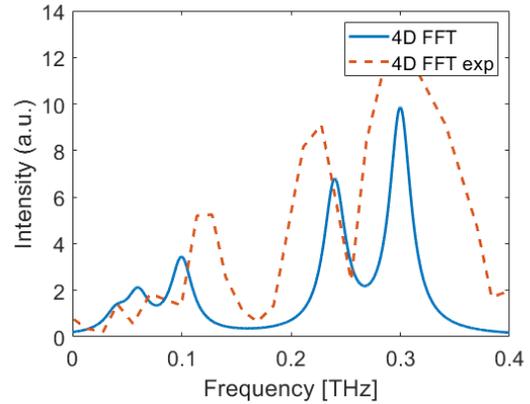

**Figure 6.** Fourier transform of the $I_{4D}$ emission function

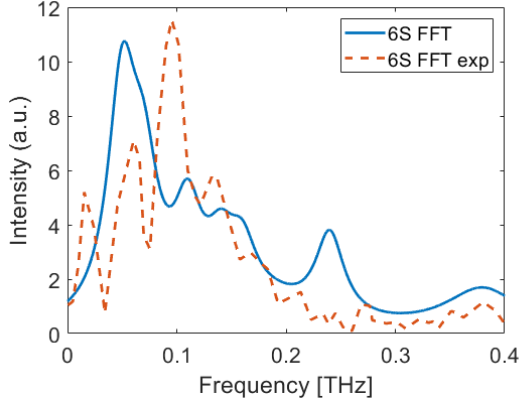

**Figure 7.** Fourier transform of the $I_{6S}$ emission function

**Table A3.** Exciton beatings characteristic frequencies of the II type
$\nu = (\tau_{qb,n_1,n_2})^{-1}$ corresponding to $(n_2 D \to n_1 S)$ transitions

| $n_2 D \downarrow / n_1 S \to$ | 3 | 4 | 5 | 6 | 7 |
|---|---|---|---|---|---|
| 3,1 | 0.368 | 0.8 | | | |
| 3,2 | 0.448 | 0.9 | | | |
| 4,1 | | | 0.207 | 0.365 | |
| 4,2 | | | 0.278 | 0.295 | |
| 5,1 | | | | 0.132 | 0.178 |
| 5,2 | | | | 0.17 | 0.134 |
| 6 | | | | 0.4 | 0.09 | 0.09 |
| 7 | | | | | 0.254 | 0.067 |

## Appendix A. Calculation details

*Appendix A.1. Exciton lifetimes*

We have calculated the exciton lifetimes using the equations (26), (29),(40,41,42) and the values of $\mu_\parallel, \mu_z$ from the Table 1, obtaining for the coefficient $\eta_{00}$, relevant for S excitons

$$\eta_{00} = 0.5 \int_0^\pi \frac{\sin\theta d\theta}{\sqrt{\sin^2\theta + (\mu_\parallel/\mu_z)_2 \cos^2\theta}} = 1.1. \quad (A.1)$$

The case od D excitons is more complicated. If the exciton S can be considered as non-degenerate, for D excitons the band structure must be taken into account. The D states reveal a splitting, where one state has a larger energy than other with the same principal quantum number. We choose two states, which result from group-theoretical considerations (for example,[13]), with representations(here for the state 4D) $4D\Gamma_3^+\Gamma_4^+(3/2)$ (state 4D1), and $4D\Gamma_5^+(5/2)$ (state D2). The symmetry properties are reflected by in the Luttinger equations for the hole effective masses (in Cu$_2$O the electron mass is assumed to be isotropic)

$$m_{hz} = \frac{m_0}{\gamma_1 - 2\gamma_2},$$
$$m_{h\parallel} = \frac{m_0}{\gamma_1 - 2\gamma_3}. \quad (A.2)$$

We take $\gamma_2 \neq \gamma_3$ for the D1 state, and $\gamma_2 = \gamma_3$ for the D2 state. Using the values $\gamma_1 = 1.79, \gamma_2 = 0.82, \gamma_3 = 0.54$ [2], we obtain $\mu_\parallel/\mu_z = 0.673$ for the D1 state and $\mu_\parallel/\mu_z = 1$ for the D2 state, and

$$\eta_{22,1} = \frac{15}{16} \int_0^\pi \frac{\sin^5\theta d\theta}{\sqrt{\sin^2\theta + 0.673\cos^2\theta}} = 1.02555,,$$

$$\eta_{22,2} = \frac{15}{16} \int_0^\pi \frac{\sin^5\theta d\theta}{\sqrt{\sin^2\theta + \cos^2\theta}} = 1. \quad (A.3)$$

The results are reported in Tables A1 and A2, and illustrated in Figures 1 and 2. For $n > 5$ we have chosen the values for D1. In Table A3 we present the values of quantum beats frequencies of the II type, calculated with the formula (20). For the lowest exciton states $3 \leq n \leq 5$ the splitting $D \to D_1, D_2$ is accounted for, thus $\Delta_{n1} = E(nD1) - E_{n00}, \Delta_{n2} = E(nD2) - E_{n00}$ and, correspondingly, $\tau_{qb,n1}, \tau_{qb,n2}$, for $n > 5$ $\Delta_{n1} = \Delta_{n2} = \Delta_n$ and $\tau_{qb,n1} = \tau_{qb,n2} = \tau_{qb,n}$, respectively. In addition, we have calculated some examples of frequencies of the I type beats (values in ps):

$$\nu_{S67} = 0.188 \quad S6 \to S7,$$
$$\nu_{D41} = 0.06 \quad D4,1 \to D4,2. \quad (A.4)$$

*Appendix A.2. The dependence on temperature and the laser power*

Since the crystal temperature is assumed to be proportional to the applied laser power $W$, we treat the exciton lifetimes shown in Table A1 as the values at $\mathcal{T} = 0$, and calculate the coherence times as functions of the applied laser power by equation (76. The results are given in Table A5. The dependence of exciton lifetimes on state number and laser power can be compared with the dependence on temperature and state number. Having in mind the experiments reported in [7], we consider the S states $n = 5, 7$ and calculate the lifetimes as function of temperature for the given states. We use the formulas

$$\tau_n(\mathcal{T}) = \mathcal{T}_0 \exp\left(n\gamma_1 - n^2\gamma_2 - \gamma_{3n}\mathcal{T}\right),$$
$$\gamma_{35} = 0.0188, \quad \gamma_{37} = 0.0376, \quad \mathcal{T}_0 \approx 3 \times 10^{-3}, \quad (A.5)$$
$$\gamma_1 = 2.592, \qquad \gamma_2 = 0.192.$$

The results are presented in Table A6. In Tables A7

**Table A1.** Exciton S ($|n00\rangle$) eigenenergies $E_{n00}$, resonances $E_{Tn00}$ (meV), and lifetimes $\tau_{n00}$ (ps), comparison of experimental data (Exp.) [7] and results of different scalings, obtained from equations (41-43)

| S | $E_{Tn00}$ | $E_{n00}$ | Exp.$\tau_{n00}$ | (41) | (42) | (43) |
|---|---|---|---|---|---|---|
| 2 | 2145.77 | 26.3 | $0.7 \pm 0.12$ | 0.88 | 0.84 | 0.96 |
| 3 | 2160.4 | 11.69 | $3.1 \pm 1$ | 2.54 | 2.52 | 3.08 |
| 4 | 2165.5 | 6.58 | $5.1 \pm 0.2$ | 5.48 | 5.51 | 6.64 |
| 5 | 2167.87 | 4.2 | $11.4 \pm 2$ | 10.08 | 10.11 | 11.17 |
| 6 | 2169.157 | 2.92 | $19 \pm 1.5$ | 16.72 | 16.61 | 15.69 |
| 7 | 2169.93 | 2.147 | $20 \pm 1.5$ | 25.76 | 25.26 | 19.24 |
| 8 | 2170.436 | 1.644 | $21.5 \pm 2$ | 37.57 | 36.32 | 21.42 |
| 9 | 2170.78 | 1.3 | $22 \pm 2.5$ | 52.53 | 50.04 | 22.43 |

**Table A2.** Exciton D ($|n22\rangle$) eigenenergies $E_{n22}$, resonances $E_{Tn22}$ (meV), and lifetimes $\tau_{n22}$ (ps) for both types of exciton D, comparison of experimental data (Exp.)[7] and results of different scalings, obtained from equations (41)-(43). The experimental values for D1, D2 excitons ($|n22\rangle_{1,2}$) from [7]

| D | $E_{Tn22}$ | $E_{n22}$ | Exp.$\tau_{n22}$ | (41) | (42) | (43) |
|---|---|---|---|---|---|---|
| 3,1 | 2161.92 | 10.29 | $3.2 \pm 0.2$ | 2.92 | 3.01 | 3.2 |
| 3,2 | 2162.42 | 10.41 | $3.2 \pm 0.2$ | 2.91 | 2.97 | 3.7 |
| 4,1 | 2166.36 | 5.86 | $5.7 \pm 0.5$ | 6.3 | 6.6 | 6.89 |
| 4,2 | 2166.65 | 5.79 | $5.7 \pm 0.5$ | 6.3 | 7.04 | 7.97 |
| 5,1 | 2168.42 | 3.75 | $10.5 \pm 0.9$ | 11.59 | 12.14 | 11.56 |
| 5,2 | 2168.6 | 3.70 | $10.5 \pm 0.9$ | 13.1 | 13.75 | 13 |
| 6 | 2169.54 | 2.57 | $17.5 \pm 1.2$ | 18.39 | 19.976 | 17.58 |
| 7 | 2170.21 | 1.89 | $19 \pm 1.5$ | 28.8 | 30.4 | 19.92 |
| 8 | 2170.65 | 1.45 | $22 \pm 2$ | 41.07 | 43 | 22.31 |
| 9 | 2170.95 | 1.14 | $23 \pm 2.5$ | 47.39 | 60.4 | 23.27 |

**Table A4.** Exciton beatings characteristic frequencies $\Omega_{n_1 n_2}$ of the II type corresponding to ($n_2 D \rightarrow n_1 S$) transitions

| $n' \downarrow / n_1 S \rightarrow$ | 3 | 4 | 5 | 6 | 7 |
|---|---|---|---|---|---|
| 3,1 | 2.31 | 5 | | | |
| 3,2 | 2.815 | | | | |
| 4,1 | | 1.3 | 2.29 | | |
| 4,2 | | 1.747 | 1.853 | | |
| 5,1 | | | 0.829 | 1.118 | |
| 5,2 | | | 1.068 | 0.842 | |
| 6 | | | 2.51 | 0.56 | 0.56 |
| 7 | | | | 1.6 | 0.42 |

**Table A5.** Lifetimes $\tau_{n00}$ vs laser power W ($\tau_{n00}$ in ps, W in mW)

| n/W | 0 | 12 | 50 | 150 |
|---|---|---|---|---|
| 3 | 2.6 | 2.4 | 1.86 | 0.96 |
| 4 | 5.23 | 4.83 | 3.76 | 1.94 |
| 5 | 9.73 | 9.0 | 7.0 | 3.61 |
| 6 | 16.7 | 15.43 | 12 | 6.2 |
| 7 | 26.45 | 24.44 | 19.0 | 9.8 |

**Table A6.** Lifetime vs temperature, $n = 5, 7(|500\rangle, |700\rangle)$, experimental data from [7], theoretical results from Eq. (78), times in ps, temperature in K

| $\mathcal{T}/\tau_n$ | Exp. $\tau_5$ | Th. | Exp. $\tau_7$ | Th. |
|---|---|---|---|---|
| 0 | | 10.49 | | 18.66 |
| 5 | $9 \pm 2$ | 9.55 | $14 \pm 2$ | 15.47 |
| 8 | $8 \pm 2$ | 9.0 | $13 \pm 2$ | 13.8 |
| 10 | $7.5 \pm 2$ | 8.7 | $15 \pm 2$ | 12.8 |
| 15 | $7.5 \pm 2$ | 7.9 | $13 \pm 2$ | 10.6 |
| 20 | $7 \pm 2$ | 7.2 | $13 \pm 2$ | 8.8 |
| 30 | $6 \pm 2$ | 5.97 | $9 \pm 2$ | 6.04 |
| 40 | $5 \pm 2$ | 4.94 | $4 \pm 2$ | 4.14 |
| 50 | $3 \pm 2$ | 4.09 | $73 \pm 2$ | 2.84 |

**Table A7.** Oscillator strengths $f_{n00}$ "pure", and results of calculations (82) (Th, in terms of $M_{01}^2$)

| States | $f_{n00}/f_{100}$ "pure" | Th. |
|---|---|---|
| $|100\rangle$ | 1 | 1 |
| $|200\rangle$ | 1/8 | 1/8.4 |
| $|300\rangle$ | 1/27 | 1/30 |
| $|400\rangle$ | 1/64 | 1/67 |
| $|500\rangle$ | 1/125 | 1/131 |
| $|600\rangle$ | 1/216 | 1/227 |
| $|700\rangle$ | 1/343 | 1/360 |
| $|800\rangle$ | 1/512 | 1/538 |

and A8 we present the results of calculations of exciton oscillator strength for S and D excitons.

*Appendix A.3. Linear and nonlinear polarization, time dependence*

The time evolution of exciton population at the given exciton state is described by the equation (99). Below we present the formulas derived from the mentioned

**Table A8.** Oscillator strengths $f_{n22}$ "pure" ((88) manuscript), and results of calculations (Results, (87) manuscript), all oscillator strength multiplied by $10^{-4} M_{02}^2 \rho_0^4$))

| States | $f_{n22}$ pure | Results |
|---|---|---|
| $|322\rangle$ | 0.81 | 0.83 |
| $|422\rangle$ | 0.48 | 0.57 |
| $|522\rangle$ | 0.28 | 0.34 |
| $|622\rangle$ | 0.18 | 0.21 |
| $|722\rangle$ | 0.12 | 0.18 |
| $|822\rangle$ | 0.08 | 0.122 |

**Table A9.** Variation of $\lambda$'s defined in equations (87) in the manuscript, with the laser power and temperature

| W | $\mathcal{T}$ | $\lambda_{th,e}$ | $\lambda_{th,hz}$ | $\lambda_{th,h\|}$ |
|---|---|---|---|---|
| 12 | 10 | 73 | 36.48 | 129 |
| 50 | 20 | 36 | 18.24 | 64.97 |
| 100 | 40 | 18.33 | 9.12 | 32.48 |
| 150 | 50 | 14.66 | 7.29 | 26.0 |

**Table A10.** Quantities $A_{500}\varphi_{500}(\rho_0), B_{500}\varphi_{500}(\rho_0), (A+B)_0 = (A_{500}\varphi_{500} + B_{500}\varphi_{500})$, laser power in mW,

| Laser power | $A_{500}\varphi_{500}$ | $B_{500}\varphi_{500}$ | $(A+B)_0$ |
|---|---|---|---|
| 12 | 27.76 | 27.45 | 55.21 |
| 50 | 16.9 | 17.6 | 34.5 |
| 100 | 3.516 | 5.16 | 8.67 |
| 150 | 1.14 | 2.59 | 3.73 |

equation for the data of 4D ($I_{4D}$) and 6S ($I_{6S}$ excitons (linear part only)

$$I_{4D} = 0.625 \exp\left(-0.11\sqrt{(t-1)^2 + 0.3}\right)$$
$$\left[1 + 0.0476\cos(0.377\,t)\right.$$
$$+ 0.32\cos(1.3\,t) + 0.3987\cos(1.76\,t) \quad \text{(A.6)}$$
$$\left. + 0.105\cos(2.29\,t) + 0.132\cos(1.85\,t)\right],$$

$$I_{6S} = \exp\left(-0.1\sqrt{(t-1)^2 + 0.3}\right)\left[1 + 0.684\cos(0.25\,t)\right.$$
$$+ 0.0074\cos(0.377\,t) + 0.148\cos(0.628\,t) \quad \text{(A.7)}$$
$$\left. + 0.296\cos(1.5\,t) + 0.44\cos(1.88\,t)\right].$$

Note that the lifetimes $\tau_n$ are taken for laser power 50 mW. The resulting line shapes are displayed in Figures 3 and 4. When the nonlinear part is accounted for, we first must determine the dependence of quantities $A_n, B_n$ on the laser power (probe temperature). We perform the calculations for the 5S state. Using the quantities $\lambda$ (Table A9) we obtain the $A_{500}, B_{500}$ values (Table A10). For the S5 state we obtain $\tau'_{05} = 9.73$, and

$$\varphi_5(\rho_0)(A_5 + B_5) = 71 \exp(-0.021\,W). \quad (A.8)$$

Separating the parameters appropriate for the state 5S, and taking $t_1 = 0.5$, we obtain

$$\chi^{(3)}_{05} = 2.745 \times 10^{-8} \left(\frac{0.5}{9.73}\right) e^{b_3 W}$$
$$\times [\varphi_5(\rho_0)(A_5 + B_5)] \left(\frac{5}{1.1}\right)^4$$
$$= W \exp(-0.0144 W) \times 3.85 \times 10^{-5}.$$

Using the function $\psi_{osc,5}(t)$ in the form ((52), manuscript)

$$\psi_{osc,5} = 713 \times \tilde{\psi}_{osc,5}(t),$$
$$\tilde{\psi}_{osc,5}(t) = 1 + 0.285 \cos(0.829\,t)$$
$$+ 0.356 \cos(1.068\,t) + 0.277 \cos(2.51\,t) \quad (A.9)$$
$$+ 0.036 \cos(2.29\,t) + 0.045 \cos(1.85\,t),$$
$$\tau_{500}(12) = 9, \quad \tau_{500}(50) = 7, \quad \tau_{500}(150) = 3.61,$$

we obtain the total normalized emission (99), for the 5S exciton, in the form

$$I_5(t,W) = N_{05} \exp\left(-\frac{|t|}{\tau_5}(W)\right) \tilde{\psi}_{osc,5}(t) \quad (A.10)$$
$$\times \left[1 - X(W) \tilde{\psi}_{osc,5}(t)\right],$$

where

$$X(W) = 2.745 \times 10^{-2} W \exp(-0.0144\,W)$$
$$\approx 0.87 \exp\left[0.035\,W - (1.925 \times 10^{-4})\,W^2\right], \quad (A.11)$$

and $N_{05}$ is a normalization constant. As can be seen from the above equation, the function $X(W)$ has a Gaussian shape, with $X(12) = 0.277$ and the maximum at $W \approx 90$. Than the function decreases to $X(150) = 0.468$.